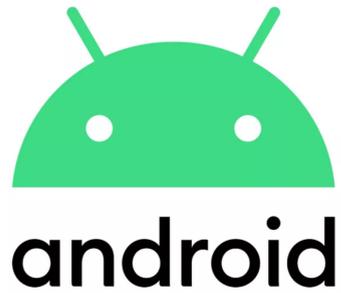

# Android OS

# Case Study

OPERATING SYSTEMS

Submitted By: -

Mayank Goel
mayank.co19@nsut.ac.in

Gourav Singal
gourav.co19@nsut.ac.in

# INDEX

# Introduction

Android is a mobile operating system based on a modified version of the Linux kernel and other open source software, designed primarily for touchscreen mobile devices such as smartphones and tablets. It is an operating system for low powered devices that run on battery and are full of hardware like Global Positioning System (GPS) receivers, cameras, light and orientation sensors, Wi-Fi and LTE (4G telephony) connectivity and a touch screen. Like all operating systems, Android enables applications to make use of the hardware features through abstraction and provide a defined environment for applications.

# History

The platform was created by Android Inc. which was bought by Google and released as the Android Open Source Project (AOSP) in 2007. Since then, Android has seen numerous updates which have incrementally improved the operating system, adding new features and fixing bugs in previous releases. Each major release is named in alphabetical order after a dessert or sugary treat, with the first few Android versions being called "Cupcake", "Donut", "Eclair", and "Froyo", in that order. During its announcement of Android KitKat in 2013, Google explained that "Since these devices make our lives so sweet, each Android version is named after a dessert". This changed after android pie.

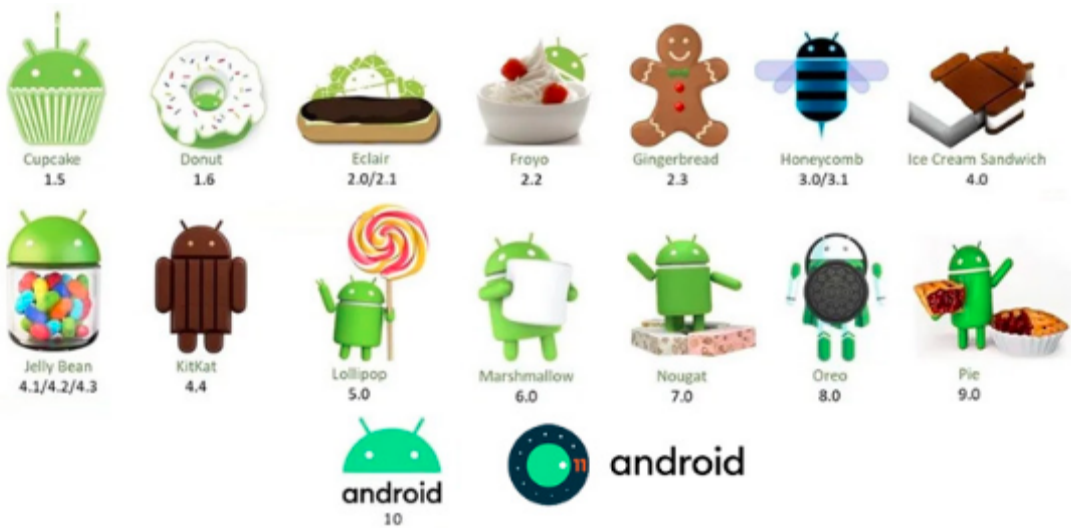

# Android Architecture

## The Linux Kernel
The foundation of the Android platform is the Linux kernel. For example, the Android Runtime (ART) relies on the Linux kernel for underlying functionalities such as threading and low-level memory management.

Using a Linux kernel allows Android to take advantage of key security features and allows device manufacturers to develop hardware drivers for a well-known kernel.

## Hardware Abstraction Layer (HAL)
The hardware abstraction layer (HAL) provides standard interfaces that expose device hardware capabilities to the higher-level Java API framework. The HAL consists of multiple library modules, each of which implements an interface for a specific type of hardware component, such as the camera or bluetooth module. When a framework API makes a call to access device hardware, the Android system loads the library module for that hardware component.

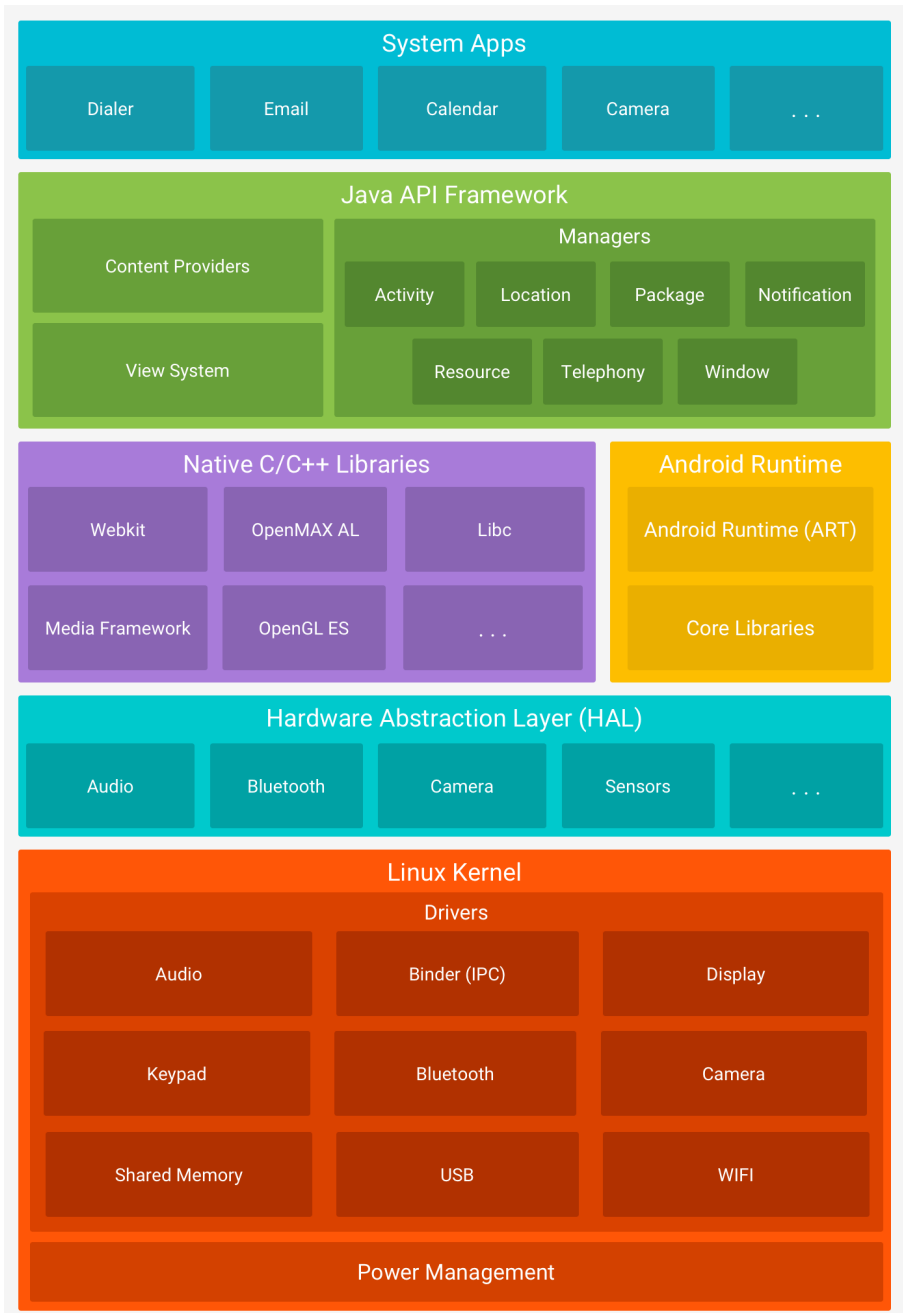

## Android Runtime
Each app runs in its own process and with its own instance of the Android Runtime (ART). ART is written to run multiple virtual machines on low-memory devices by executing DEX files, a bytecode format designed specially for Android that's optimized for minimal memory footprint. Build tools, such as d8, compile Java sources into DEX bytecode, which can run on the Android platform.

# Native C/C++ Libraries

Many core Android system components and services, such as ART and HAL, are built from native code that require native libraries written in C and C++. The Android platform provides Java framework APIs to expose the functionality of some of these native libraries to apps. For example, you can access OpenGL ES through the Android framework's Java OpenGL API to add support for drawing and manipulating 2D and 3D graphics in your app.

# Java API Framework

The entire feature-set of the Android OS is available to you through APIs written in the Java language. These APIs form the building blocks you need to create Android apps by simplifying the reuse of core, modular system components and services, which include the following:

- A rich and extensible View System you can use to build an app's UI, including lists, grids, text boxes, buttons, and even an embeddable web browser
- A Resource Manager, providing access to non-code resources such as localized strings, graphics, and layout files
- A Notification Manager that enables all apps to display custom alerts in the status bar
- An Activity Manager that manages the lifecycle of apps and provides a common navigation back stack
- Content Providers that enable apps to access data from other apps, such as the Contacts app, or to share their own data

# System Apps

Android comes with a set of core apps for email, SMS messaging, calendars, internet browsing, contacts, and more. Apps included with the platform have no special status among the apps the user chooses to install. So a third-party app can become the user's default web browser, SMS messenger, or even the default keyboard (some exceptions apply, such as the system's Settings app).

# Kernel and Startup

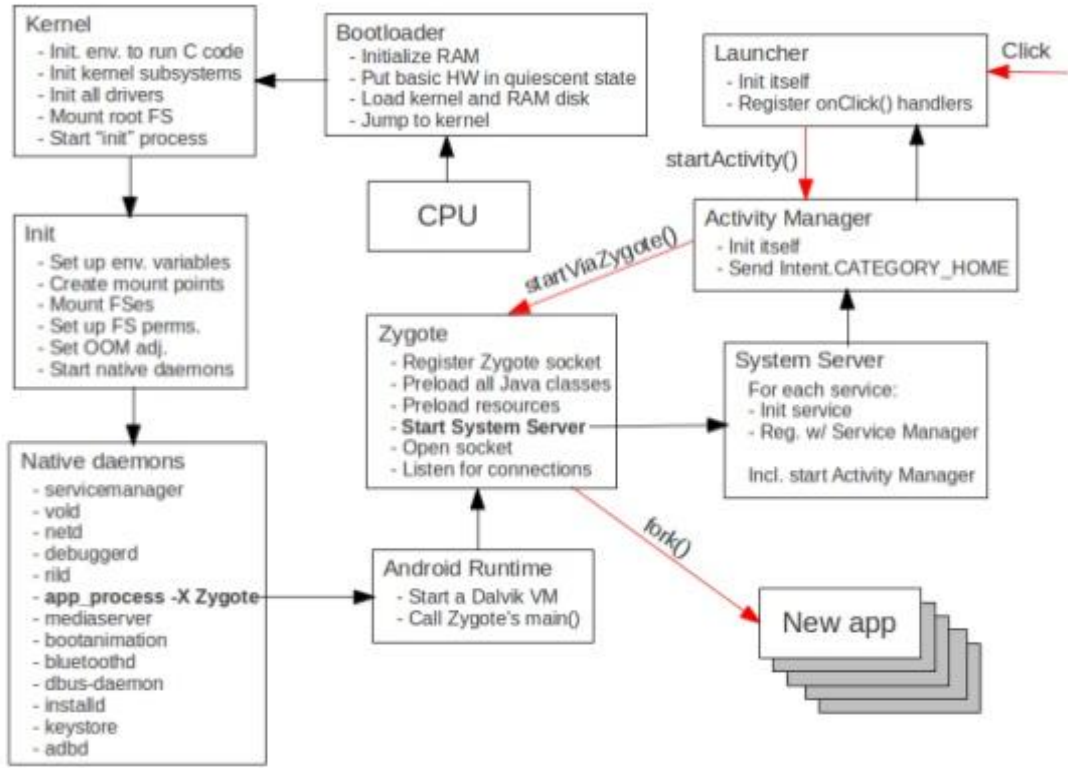

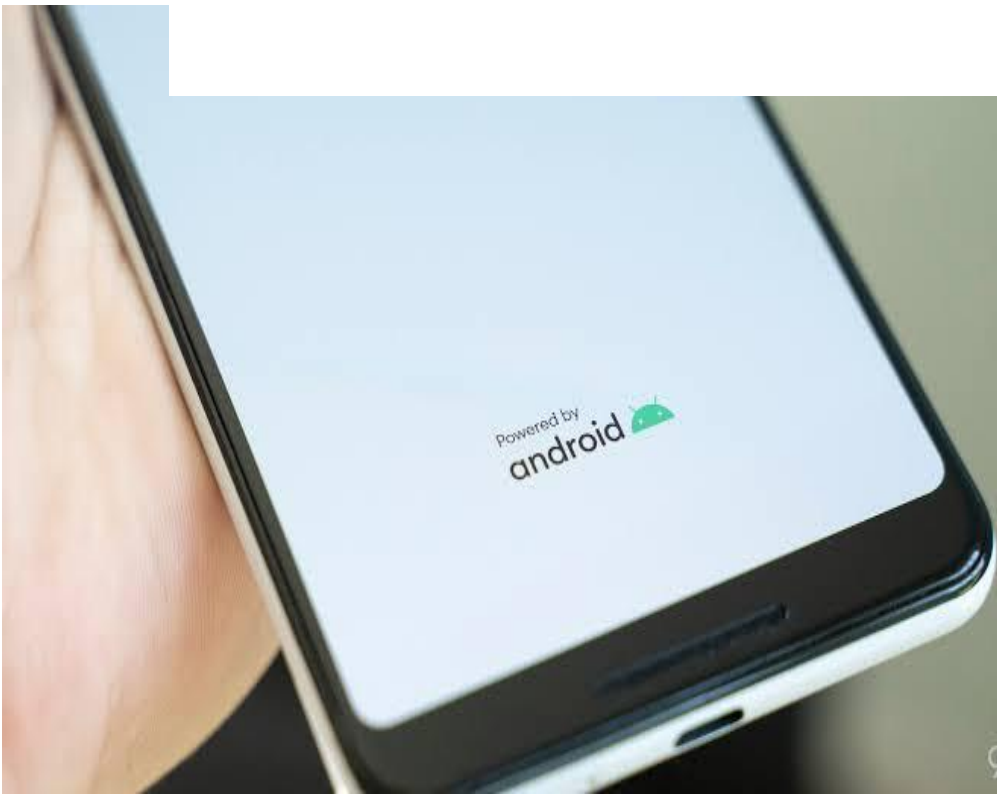

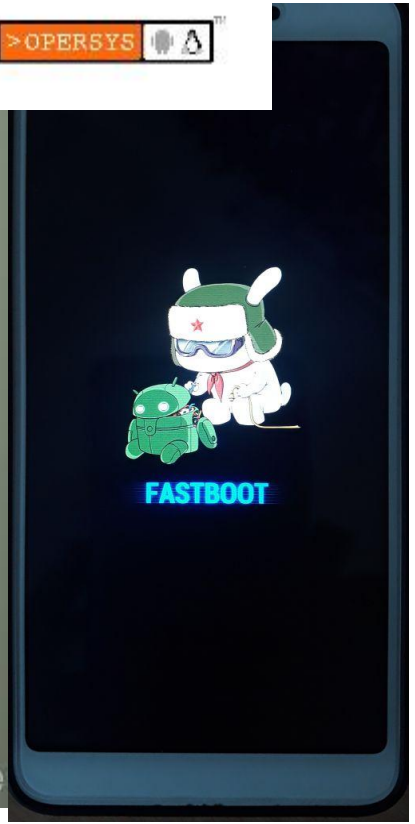

# Process Management

## The Process Lifecycle Hierarchy

A process on Android can be in one of five different states at any given time, from most important to least important:

1. Foreground process: The app you're using is considered the foreground process. Other processes can also be considered foreground processes.
2. Visible process: A visible process isn't in the foreground, but is still affecting what you see on your screen.
3. Service process: A service process isn't tied to any app that's visible on your screen. However, it's doing something in the background.
4. Background process: Background processes are not currently visible to the user. They have no impact on the experience of using the phone. At any given time, many background processes are currently running. They're kept in memory so you can quickly resume using them when you go back to them, but they aren't using valuable CPU time or other non-memory resources.
5. Empty process: An empty process doesn't contain any app data anymore. It may be kept around for caching purposes to speed up app launches later, or the system may kill it as necessary.

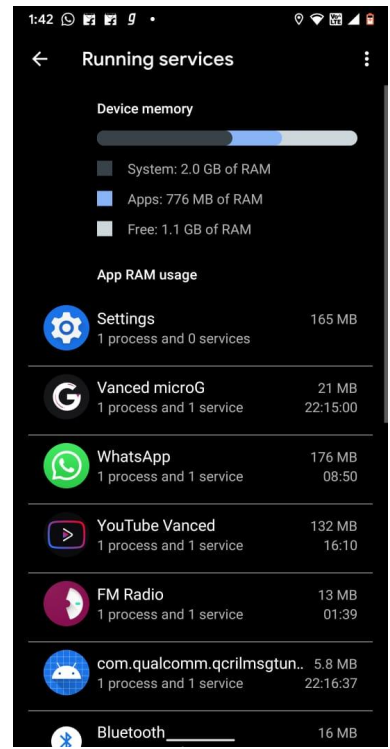

Android supports both manual and automatic process management.

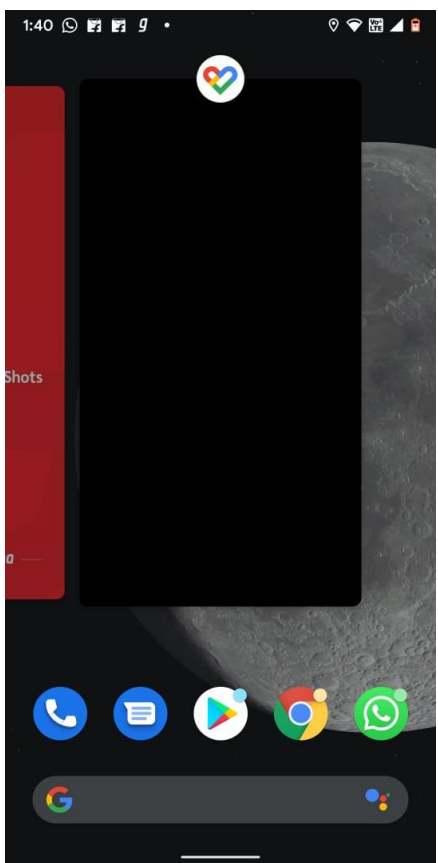
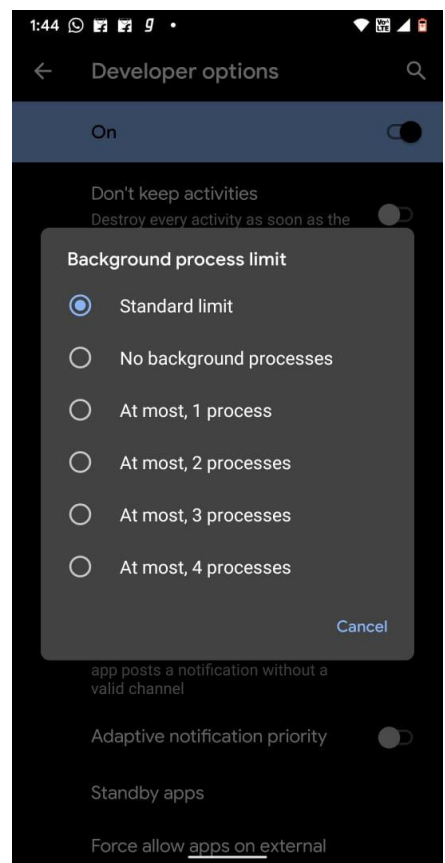

First figure shows how we ourselves manage apps, how we kill them and save our memory. In the second figure , we see how the OS automatically sets a limit on background processes but it can be easily modified.

## Threads

A thread is a thread of execution in a program. The Java Virtual Machine allows an application to have multiple threads of execution running concurrently.

Every thread has a priority. Threads with higher priority are executed in preference to threads with lower priority. Each thread may or may not also be marked as a daemon. When code running in some thread creates a new Thread object, the new thread has its priority initially set equal to the priority of the creating thread, and is a daemon thread if and only if the creating thread is a daemon.

## Inter Process Communication (IPC)

As android is based on linux OS we can use traditional Linux techniques such as network sockets and shared files. Android system functionality for IPC are Intent, Binder or Messenger with a Service, and BroadcastReceiver. The Android IPC mechanisms allow us to verify the identity of the application connecting to your IPC and set security policy for each IPC mechanism.

There are two ways processes communicate with one another:

1. Using Intents:

    An intent is to perform an action on the screen. It is mostly used to start an activity, send a broadcast receiver,start services and send messages between two activities.

    Intent intent = new Intent(this, SomeActivity.class); intent.putExtra("key", "some serialisable data"); *SENDING MESSAGE* startActivity(intent);

2. Using Android Interface Definition Language (AIDL)

    It allows us to define the programming interface that both the client and service agree upon in order to communicate with each other using interprocess communication (IPC). On Android, one process cannot normally access the memory of another process. So to talk, they need to decompose their objects into primitives that the operating system can understand, and marshall the objects across that boundary for you.

# How is DeadLock Handled?

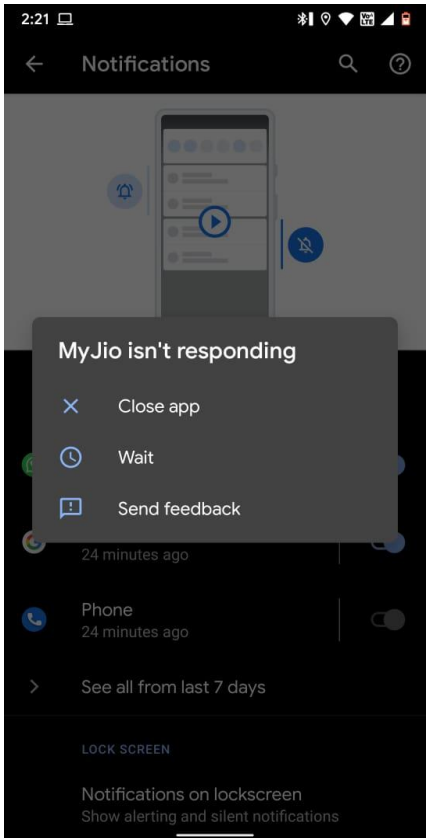

Deadlock is a common platform in multiprocessing systems.

And there are 3 ways most of the OS used to deal with deadlock they are as follow:

1) Prevent or Avoid
2) Detect and recover
3) Ignore the problem

Android uses a second option to handle the deadlocks, as it is hard to prevent from occuring deadlocks it let deadlocks to occur. Then the system detects that a deadlock has occurred and tries to recover from it. Detecting deadlocks is easily possible since each process has locked and is waiting for one another. Now it has to recover, an algorithm has to track resource allocation and process states, it goes back and starts over again with one or more of the processes in order to remove the deadlock.

There are majorly 4 conditions which if get satisfied then we can prevent deadlock

Conditions are

1) Mutual Exclusion
2) Hold And Wait
3) No Preemption
4) Circular Wait

Deadlock avoidance can be done with Banker's Algorithm

1. Banker's Algorithm

   It is a resource allocation and deadlock avoidance algorithm which tests all the requests made by processes for resources, it checks for the safe state, if after granting request system remains in the safe state it allows the request and if there is no safe state it does not allow the request made by the process.

2. Preventing recursive locks

# CPU Scheduling

The Android operating system uses O(1) scheduling algorithm as it is based on Linux Kernel 2.6. Therefore the scheduler is named as Completely Fair Scheduler as the processes can schedule within a constant amount of time, regardless of how many processes are running on the operating system.

The process "nice" level impacts the normal "fair" scheduling policy of Linux; threads that have a higher niceness will be run less often than threads with a lower niceness hus if you set a thread's priority to Process.THREAD_PRIORITY_BACKGROUND or greater, you will also be putting the thread into the background cgroup. Set it to Process.THREAD_PRIORITY_DEFAULT and it will be in the default cgroup.. In the situation where you have one thread at a "default" priority (as defined in Process.THREAD_PRIORITY_DEFAULT) will get to run significantly more often than those at a background priority (or Process.THREAD_PRIORITY_BACKGROUND).

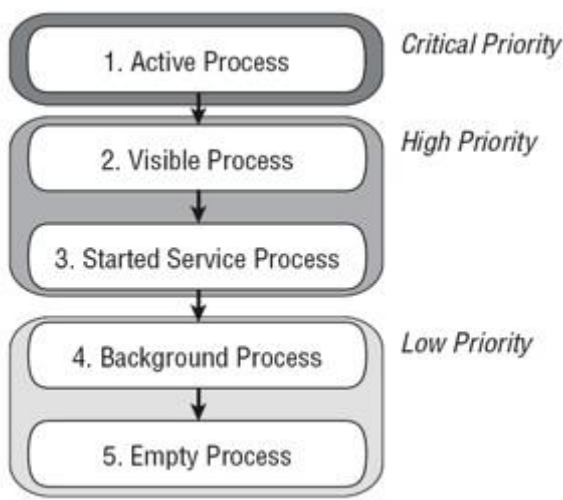

In theory this can make sure that the foreground/UI threads aren't impacted significantly by background work… however, in practice, it isn't sufficient.
To address this, Android also uses Linux cgroups in a simple way to create more strict foreground vs. background scheduling. The foreground/default cgroup allows thread scheduling as normal. The background cgroup however applies a limit of only some small percent of the total CPU time being available to all threads in that cgroup. Android implicitly moves threads between the default and background cgroups when you use its public APIs to set the thread priority.

Because of this, by following the normal convention of putting your background worker threads into the background priority you can ensure that they do not disrupt your foreground UI thread.

In addition, Android will also move all threads in a process to the background cgroup for processes that it knows are not critical to the user. Any background process or service process has its threads put into the background cgroup, regardless of whether individual threads have requested a foreground scheduling priority.

# Memory Management

The Android Runtime and Dalvik virtual machine use paging and memory-mapping to manage memory. This means that any memory an app modifies-whether by allocating new objects or touching mmapped pages-remains resident in RAM and cannot be paged out

The only way to release memory from an app is to release object references that the app holds, making the memory available to the garbage collector. That is with one exception: any files mmapped in without modification, such as code, can be paged out of RAM if the system wants to use that memory elsewhere

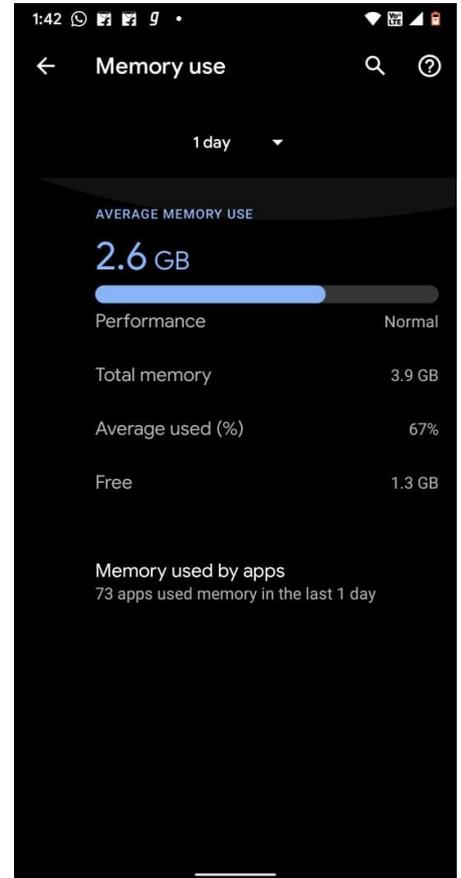

Garbage Collection

A managed memory environment, like the ART or Dalvik virtual machine, keeps track of each memory allocation. Once it determines that a piece of memory is no longer being used by the program, it frees it back to the heap, without any intervention from the programmer. The mechanism for reclaiming unused memory within a managed memory environment is known as garbage collection. Garbage collection has two goals: find data objects in a program that cannot be accessed in the future; and reclaim the resources used by those objects.

Share Memory

In order to fit everything it needs in RAM, Android tries to share RAM pages across processes. It can do so in the following ways:

1) Each app process is forked from an existing process called Zygote. The Zygote process starts when the system boots and loads common framework code and resources (such as activity themes).

2) Most static data is mapped into a process. This technique allows data to be shared between processes, and also allows it to be paged out when needed

3) In many places, Android shares the same dynamic RAM across processes using explicitly allocated shared memory regions (either with ashmem or gralloc)

Allocation and reclaim app memory

The Dalvik heap is constrained to a single virtual memory range for each app process. This defines the logical heap size, which can grow as it needs to but only up to a limit that the system defines for each app.

Restrict app Memory

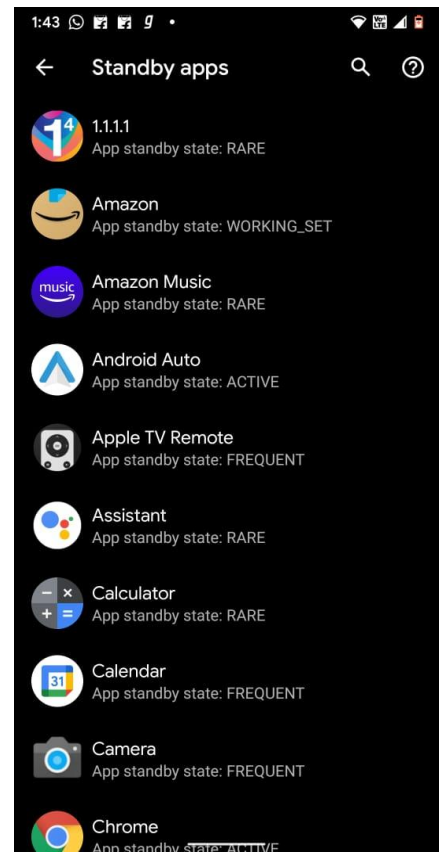

To maintain a functional multi-tasking environment, Android sets a hard limit on the heap size for each app. The exact heap size limit varies between devices based on how much RAM the device has available overall. If your app has reached the heap capacity and tries to allocate more memory, it can receive an OutOfMemoryError.

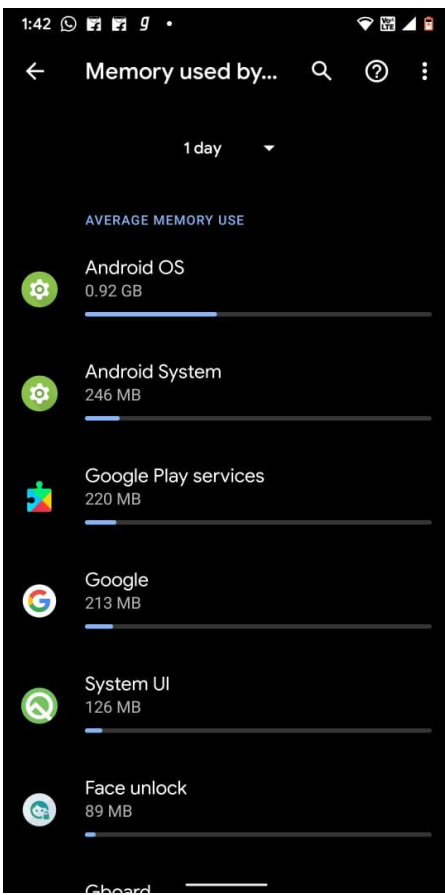

Switch Apps

When users switch between apps, Android keeps apps that are not foreground—that is, not visible to the user or running a foreground service like music playback— in a cache. For example, when a user first launches an app, a process is created for it; but when the user leaves the app, that process does not quit. The system keeps the process cached. If the user later returns to the app, the system reuses the process, thereby making the app switching faster.

# Storage Management

There are two things that the OS is supposed to do in managing the Internal storage.

First lookout for filled space and others give app access to read/write in internal storage. In android 10 and above apps can be given only access to media elements of the device. Most android use e-MMC based storage. They provide upto 300MBps read and write speed.

File system in android is that of a linux having a root directory, they can be seen in the image on the right. There are various partitions each its features. Only internal storage or memory cards all are managed by OS 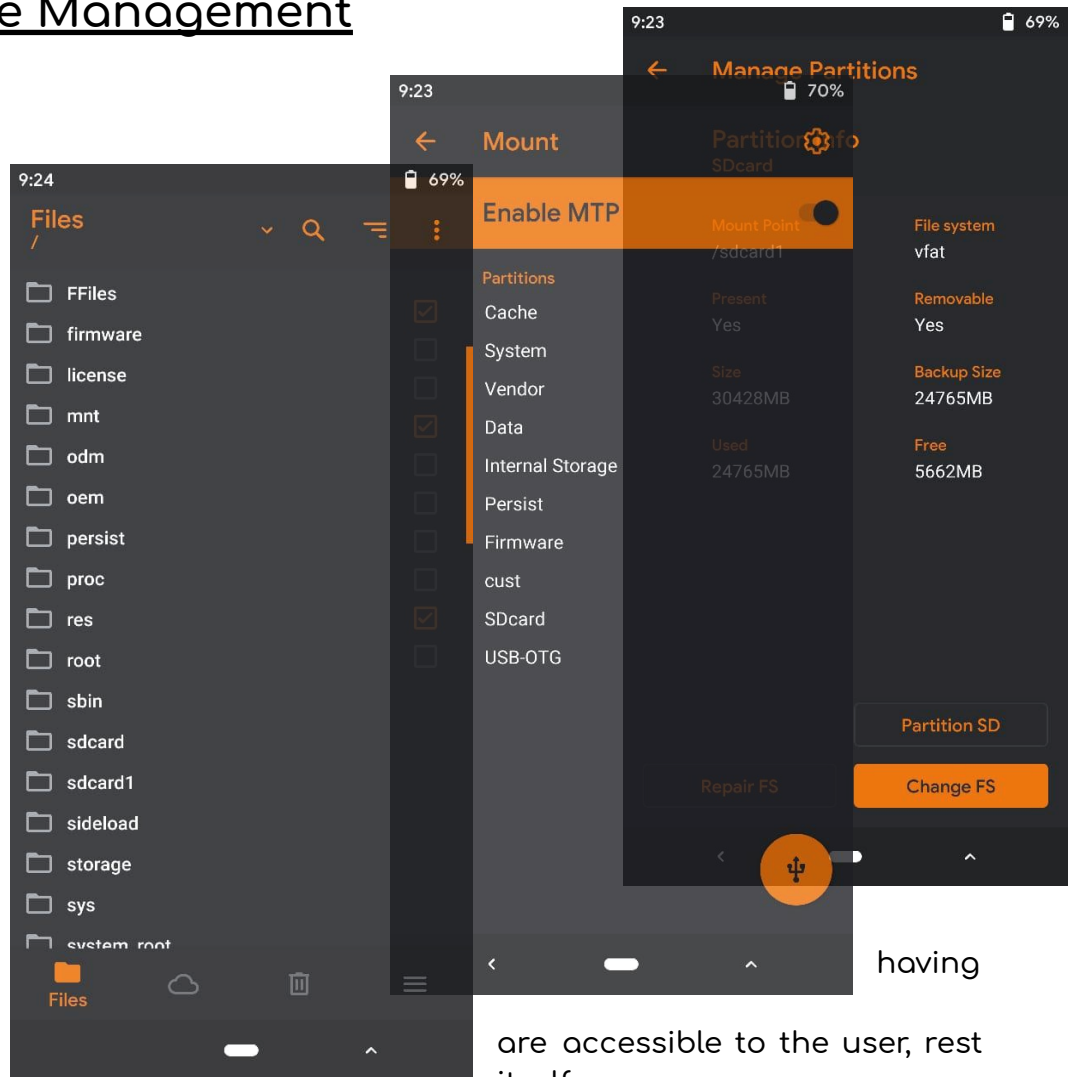 having are accessible to the user, rest itself.

StorageManager is the interface to the system's storage service. The storage manager handles storage-related items such as Opaque Binary Blobs (OBBs).

OBBs contain a filesystem that can be encrypted on disk and mounted on-demand from an application. OBBs are a good way of providing large amounts of binary assets without packaging them into APKs as they may be multiple gigabytes in size. However, due to their size, they're most likely stored in a shared storage pool accessible from all programs. The system does not guarantee the security of the OBB file itself: if any program modifies the OBB, there is no guarantee that a read from that OBB will produce the expected output.

# I/O Management

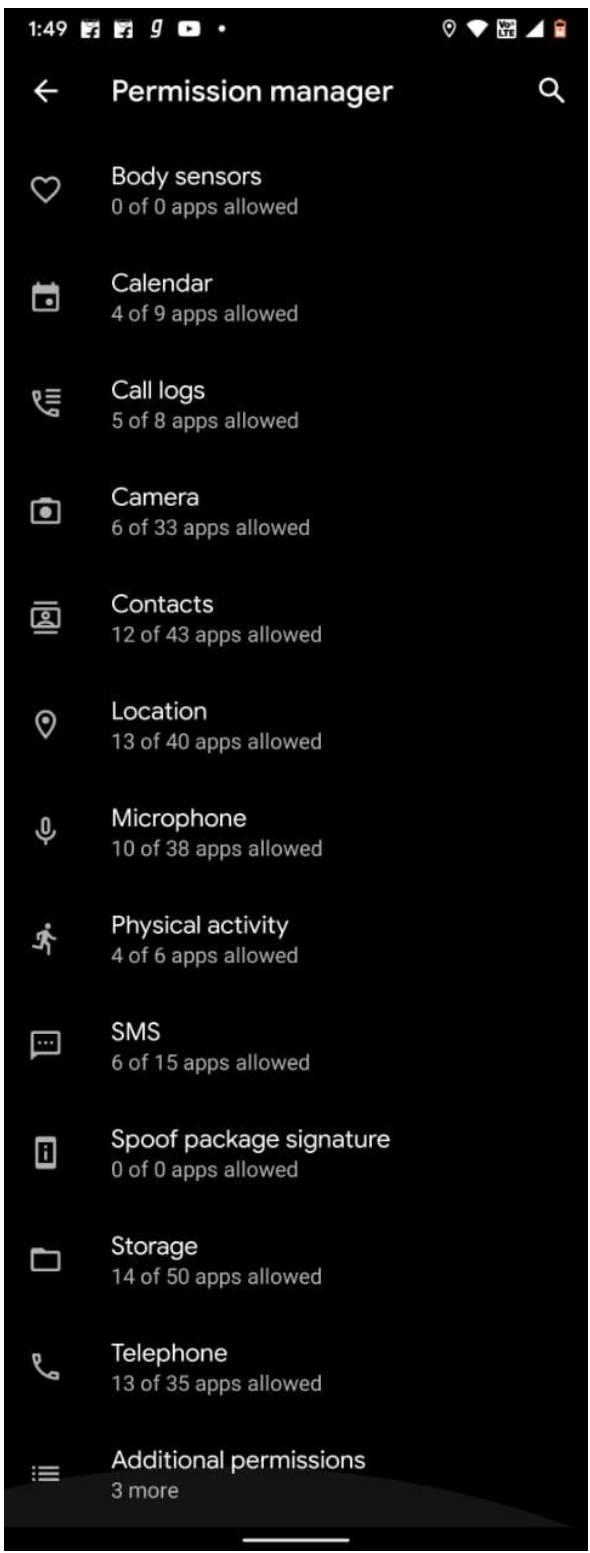

One of the important jobs of an Operating System is to manage various I/O devices including mouse, keyboards, touch pad, disk drives, display adapters, USB devices, Bit-mapped screen, LED, Analog-to-digital converter, On/off switch, network connections, audio I/O, printers etc.

An I/O system is required to take an application I/O request and send it to the physical device, then take whatever response comes back from the device and send it to the application. I/O devices can be divided into two categories –

  1) Block devices – A block device is one with which the driver communicates by sending entire blocks of data. For example, Hard disks, USB cameras, Disk-On-Key etc.

  2) Character devices – A character device is one with which the driver communicates by sending and receiving single characters (bytes, octets). For example, serial ports, parallel ports, sounds cards etc

## Device Controllers

Device drivers are software modules that can be plugged into an OS to handle a particular device. The Operating System takes help from device drivers to handle all I/O devices.

The Device Controller works like an interface between a device and a device driver. I/O units (Keyboard, mouse, printer, etc.) typically consist of a mechanical component and an electronic component where the electronic component is called the device controller.

## Synchronous vs asynchronous I/O

Synchronous I/O – In this scheme CPU execution waits while I/O proceeds

Asynchronous I/O – I/O proceeds concurrently with CPU execution

## Communication to I/O Devices

The CPU must have a way to pass information to and from an I/O device. There are three approaches available to communicate with the CPU and Device.

- Special Instruction I/O

This uses CPU instructions that are specifically made for controlling I/O devices. These instructions typically allow data to be sent to an I/O device or read from an I/O device.

- Memory-mapped I/O

When using memory-mapped I/O, the same address space is shared by memory and I/O devices. The device is connected directly to certain main memory locations so that I/O devices can transfer blocks of data to/from memory without going through the CPU.

- Direct memory access (DMA)

Slow devices like keyboards will generate an interrupt to the main CPU after each byte is transferred. If a fast device such as a disk generated an interrupt for each byte, the operating system would spend most of its time handling these interrupts. So a typical computer uses direct memory access (DMA) hardware to reduce this overhead.

Direct Memory Access (DMA) means CPU grants I/O module authority to read from or write to memory without involvement. The DMA module itself controls exchange of data between main memory and the I/O device. The CPU is only involved at the beginning and end of the transfer and interrupted only after the entire block has been transferred.

## Polling vs Interrupts I/O

A computer must have a way of detecting the arrival of any type of input. There are two ways that this can happen, known as polling and interrupts. Both of these techniques allow the processor to deal with events that can happen at any time and that are not related to the process it is currently running.

## Interrupts I/O

An alternative scheme for dealing with I/O is the interrupt-driven method. An interrupt is a signal to the microprocessor from a device that requires attention.

A device controller puts an interrupt signal on the bus when it needs CPU's attention when CPU receives an interrupt, It saves its current state and invokes the appropriate interrupt handler using the interrupt vector (addresses of OS routines to handle various events). When the interrupting device has been dealt with, the CPU continues with its original task as if it had never been interrupted.

I/O software is often organized in the following layers –

1. User Level Libraries – This provides a simple interface to the user program to perform input and output. For example, stdio is a library provided by C and C++ programming languages.

2. Kernel Level Modules – This provides device drivers to interact with the device controller and device independent I/O modules used by the device drivers.

3. Hardware – This layer includes actual hardware and hardware controllers which interact with the device drivers and makes hardware alive.

A key concept in the design of I/O software is that it should be device independent where it should be possible to write programs that can access any I/O device without having to specify the device in advance. For example, a program that reads a file as input should be able to read a file on a floppy disk, on a hard disk, or on a CD-ROM, without having to modify the program for each different device.

## Device Drivers

Device drivers are software modules that can be plugged into an OS to handle a particular device. The Operating System takes help from device drivers to handle all I/O devices.

A device driver performs the following jobs –

1) To accept requests from the device independent software above to it.
2) Interact with the device controller to take and give I/O and perform required error handling
3) Making sure that the request is executed successfully

## Interrupt handlers

An interrupt handler, also known as an interrupt service routine or ISR, is a piece of software or more specifically a callback function in an operating system or more specifically in a device driver, whose execution is triggered by the reception of an interrupt.

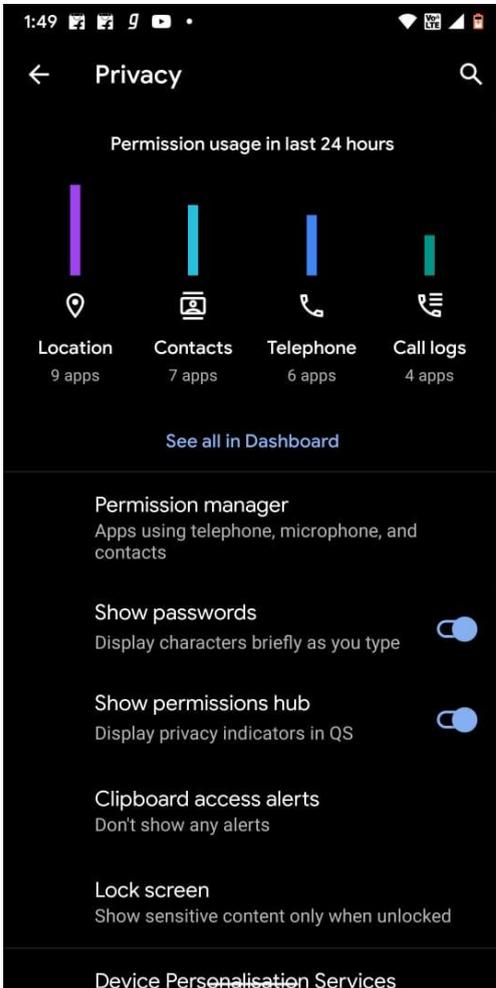

## Device-Independent I/O Software

The basic function of the device-independent software is to perform the I/O functions that are common to all devices and to provide a uniform interface to the user-level software. Though it is difficult to write completely device independent software, we can write some modules which are common among all the devices. Following is a list of functions of device-independent I/O Software –

1. Uniform interfacing for device drivers
2. Device naming - Mnemonic names mapped to Major and Minor device numbers
3. Device protection
4. Providing a device-independent block size
5. Buffering because data coming off a device cannot be stored in the final destination.
6. Storage allocation on block devices
7. Allocation and releasing dedicated devices
8. Error Reporting

## User-Space I/O Software

These are the libraries which provide a richer and simplified interface to access the functionality of the kernel or ultimately interactive with the device drivers. Most of the user-level I/O software consists of library procedures with some exceptions like spooling system which is a way of dealing with dedicated I/O devices in a multiprogramming system.

## Kernel I/O Subsystem

Kernel I/O Subsystem is responsible to provide many services related to I/O. Following are some of the services provided.

- **Scheduling** – Kernel schedules a set of I/O requests to determine a good order in which to execute them. When an application issues a blocking I/O system call, the request is placed on the queue for that device. The Kernel I/O scheduler rearranges the order of the queue to improve the overall system efficiency and the average response time experienced by the applications.

- **Buffering** – Kernel I/O Subsystem maintains a memory area known as a buffer that stores data while they are transferred between two devices or between devices with an application operation. Buffering is done to cope with a speed mismatch between the producer and consumer of a data stream or to adapt between devices that have different data transfer sizes.

- **Caching** – Kernel maintains cache memory which is a region of fast memory that holds copies of data. Access to the cached copy is more efficient than access to the original.

- **Spooling and Device Reservation** – A spool is a buffer that holds output for a device, such as a printer, that cannot accept interleaved data streams. The spooling system copies the queued spool files to the printer one at a time. In some operating systems, spooling is managed by a system daemon process. In other operating systems, it is handled by an in kernel thread.

- **Error Handling** – An operating system that uses protected memory can guard against many kinds of hardware and application errors.
- **Main memory and the I/O device** - The CPU is only involved at the beginning and end of the transfer and interrupted only after the entire block has been transferred.

# Battery Optimization

Battery life is the single most important aspect of the mobile user experience. A device without power offers no functionality at all. For this reason, it is critically important that apps be as respectful of battery life as possible.

There are three important things to keep in mind in keeping your app power-thrifty:

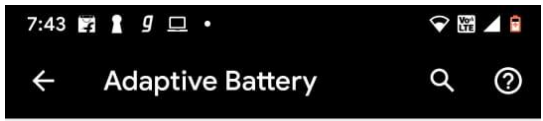

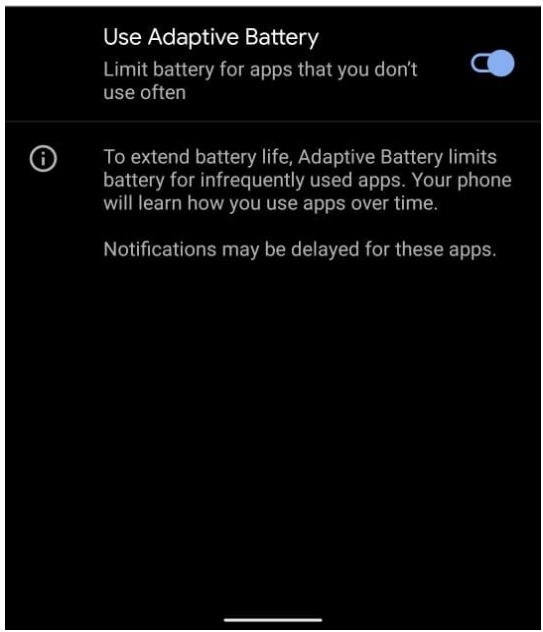

3) Make your apps Lazy First.
4) Take advantage of platform features that can help manage your app's battery consumption.
5) Use tools that can help you identify battery-draining culprits.

## Lazy First

Making your app Lazy First means looking for ways to reduce and optimize operations that are particularly battery-intensive. The core questions underpinning Lazy First design are:

1) Reduce: Are there redundant operations your app can cut out?
2) Defer: Does an app need to perform an action right away?
3) Coalesce: Can work be batched, instead of putting the device into an active state many times?

## Platform features

Broadly speaking, the Android platform provides two categories of help for you to optimize your app's battery use. First, it provides several APIs that you can implement in your app.

There are also internal mechanisms in the platform to help conserve battery life. While they are not APIs that you implement programmatically, you should still be aware of them so that your app can leverage them successfully.

Additionally, Android 9 (API level 28) makes a number of improvements to battery saver mode. Device manufacturers determine the precise restrictions imposed. As an example, on AOSP builds, the system applies the following restrictions:

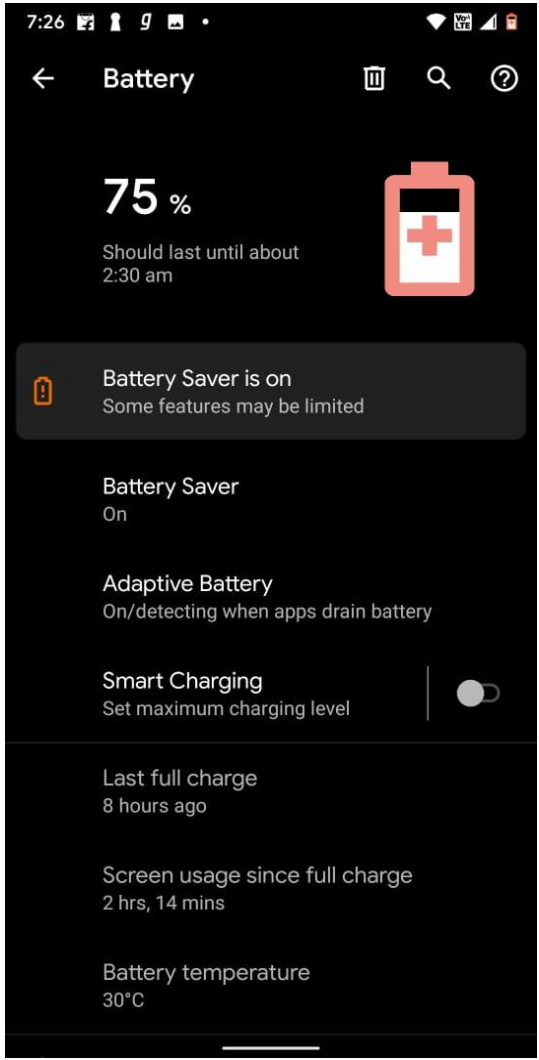

1) The system puts apps in app standby mode more aggressively, instead of waiting for the app to be idle.
2) Background execution limits apply to all apps, regardless of their target API level.
3) Location services may be disabled when the screen is off.
4) Background apps do not have network access.

As always, it's a good idea to test your app while the battery saver is active. You can turn on the battery saver manually through the device's Settings > Battery Saver screen.

## Tooling

You can get even more benefit out of these features by using the tools available for the platform to discover the parts of your app that consume the most power. Finding what to target is a big step toward successful optimization.

There are tools for Android, including Profile GPU Rendering and Battery Historian to help you identify areas that you can optimize for better battery life. Take advantage of these tools to target areas where you can apply the principles of Lazy First.

# References

- https://developer.android.com/guide/platform

- https://en.wikipedia.org/wiki/Android_(operating_system)

- https://en.wikipedia.org/wiki/Dalvik_(software)#:~:text=Dalvik%20was%20an%20integral%20part%20of%20the%20Android,smart%20TVs%20and%20wearables.%20Dalvik%20is%20open-source%20software%2C

- https://www.howtogeek.com/161225/htg-explains-how-android-manages-processes/#:~:text=%20How%20Android%20Manages%20Processes%20%201%20The,do%20it%20if%20you%20want.%20You…%20More%20

- https://prezi.com/pwxtdbdndbym/introduction-of-android-os/#:~:text=Deadlock%20is%20a%20common%20problem%20in%20multiprocessing%20systems.&text=Android%20use%20second%20option%20to,it%20let%20deadlocks%20to%20occur.&text=Detecting%20a%20deadlocks%20is%20easily,are%20waiting%20for%20one%20another

- https://1drv.ms/w/s!AjpNOC42EnTzgZ51WlbzEw80-kIJEg?e=KHnxJc

- https://www.researchgate.net/profile/Ahamed-Shibly/publication/299394606_Android_Operating_System_Architecture_Security_Challenges_and_Solutions/links/56f3da5e08ae81582bef1a0b/Android-Operating-System-Architecture-Security-Challenges-and-Solutions.pdf?origin=publication_detail

- Operating System Concepts *by* Avi Silberschatz, Greg Gagne, and Peter Baer Galvin